\documentclass[sigconf]{acmart}
\usepackage{enumitem}
\usepackage{subcaption}
\usepackage{hyperref}

\AtBeginDocument{%
  }

\acmConference[Under Review]{Submitted to WebScience 2025}

\begin{document}

\title{Ketto and the Science of Giving: \\ A Data-Driven Investigation of Crowdfunding for India}

\author{Karuna Chandra}
 \authornote{Both authors contributed equally to this research.}
\affiliation{%
\institution{IIIT-Hyderabad}\country{India}}
\email{zelekb@rpi.edu}

\author{Akshay Menon}
\authornotemark[1]
\affiliation{%
\institution{IIIT-Hyderabad}\country{India}}
\email{akshaymjry@gmail.com}

\author{Lydia Manikonda}
\affiliation{%
\institution{Rensselaer Polytechnic Institute}\country{USA}}
\authornote{Corresponding author.}
\email{manikl@rpi.edu}

\author{Ponnurangam Kumaraguru}
\affiliation{%
\institution{IIIT-Hyderabad}\country{India}}
\email{pk.guru@iiit.ac.in}









\begin{abstract}

The main goal of this paper is to investigate an up and coming crowdfunding platform used to raise funds for social causes in India called Ketto. Despite the growing usage of this platform, there is insufficient understanding in terms of why users choose this platform when there are other popular platforms such as GoFundMe. Using a dataset comprising of 119,493 Ketto campaigns, our research conducts an in-depth investigation into different aspects of how the campaigns on Ketto work with a specific focus on medical campaigns, which make up the largest percentage of social causes in the dataset. We also perform predictive modeling to identify the factors that contribute to the success of campaigns on this platform. We use several features such as the campaign metadata, description, geolocation, donor behaviors, and campaign-related features to learn about the platform and its components. Our results suggest that majority of the campaigns for medical causes seek funds to address chronic health conditions, yet medical campaigns have the least success rate. Most of the campaigns originate from the most populous states and major metropolitan cities in India. Our analysis also indicates that factors such as online engagement on the platform in terms of the number of comments, duration of the campaign, and frequent updates on a campaign positively influence the funds being raised. Overall, this preliminary work sheds light on the importance of investigating various dynamics around crowdfunding for India-focused community-driven needs.  
\end{abstract}



\keywords{Medical Crowdfunding, Web Platforms, Quantitative Analysis}


\maketitle

\section{Introduction}

\begin{figure}
    \centering
    \includegraphics[width=\linewidth]{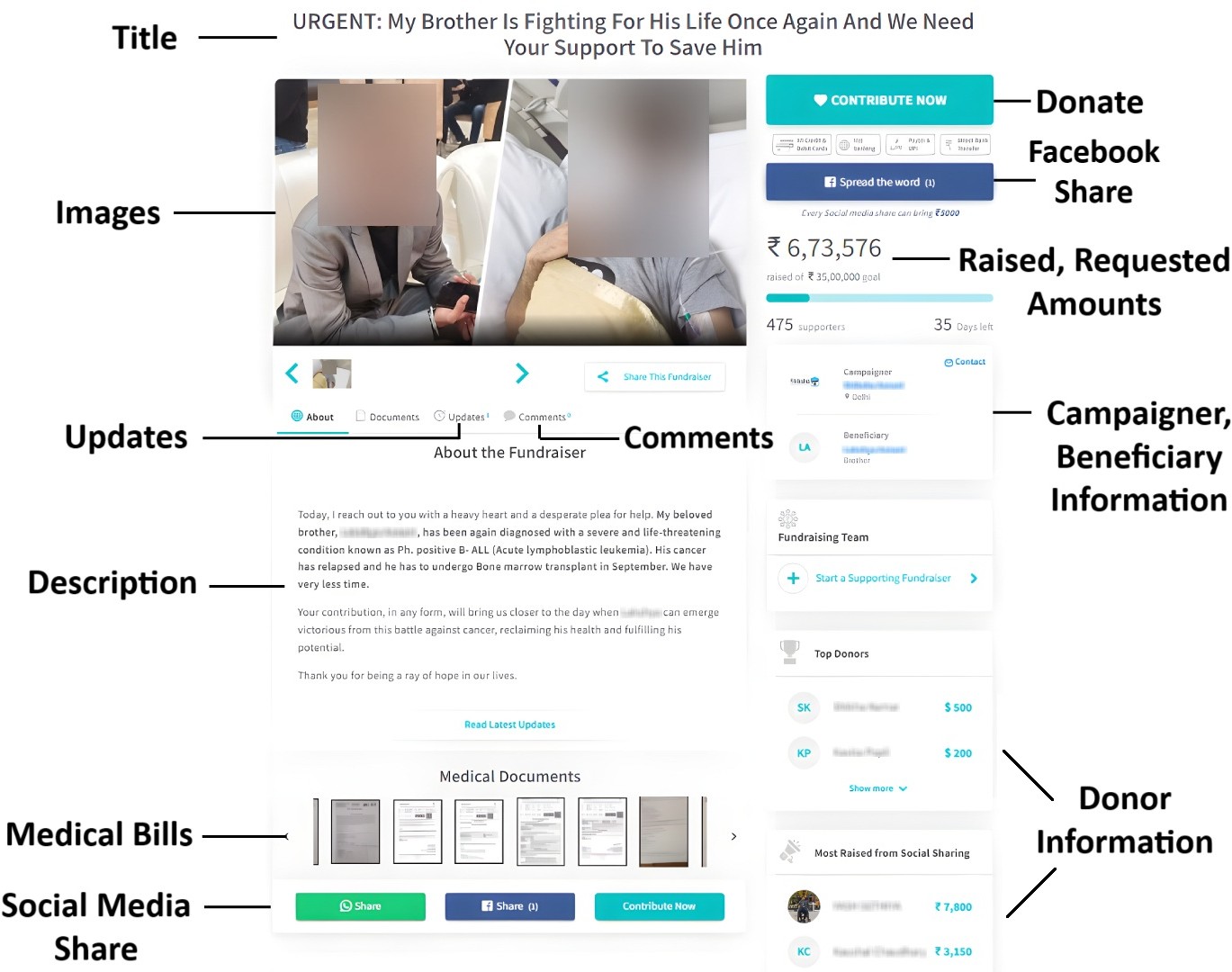}
    \caption{Example of a Ketto campaign, showing different metadata elements such as images, campaign description, documents including medical bills, updates and comments sections, top donors, and options to share the campaign through social media platforms.}
    \label{fig:SampleKettoCampaign}
\end{figure}

Ketto~\cite{KettoWebsite} is a prominent Indian charity-based digital crowdfunding platform that facilitates user engagement to support various social causes via campaigns mostly for people in the Indian subcontinent. Despite its growing popularity, there is limited understanding on the main purpose and dynamics of Ketto, specifically regarding why people use the platform and in what contexts. The uniqueness of this platform with regards to helping beneficiaries with varying cultural backgrounds may shed light on the social iniquities and potential resource-related gaps. Since its inception in 2012, Ketto has facilitated over 320,000 campaigns. Each campaign typically involves a beneficiary---the individual in need of funds---and a campaigner who manages the process. These initiatives span several categories, including medical, children, education, and memorial causes. Figure~\ref{fig:SampleKettoCampaign} shows a campaign on Ketto and the metadata elements associated with it. Unlike many popular crowdfunding platforms that employ an `all or nothing' model like \url{Kickstarter.com}, Ketto allows campaigners to withdraw raised funds even if the goal amount is not achieved. This feature is particularly beneficial for large target amounts where full attainment is less likely. The platform also enables likes and comments on campaigns and multiple updates from campaigners. Thus, we aim to answer the following research questions in this paper: 1) \emph{What is Ketto and what types of social causes are supported on the platform?} 2) \emph{When it comes to medical crowdfunding campaigns, what are the differences between campaigns in terms of the posted content? } and 3) \emph{What factors contribute to the success of campaigns on Ketto?}

India faces significant public health challenges with high rates of diseases and a large percentage of underserved populations. Thus, studying this platform is a vital contribution to help shed light on healthcare accessibility. While Ketto is a platform mainly used in the Indian subcontinent, it provides a more contained look into online crowdfunding as a whole. By familiarizing audiences with Ketto and its associated context, conclusions about the success of campaigns can be made and applied specifically to these Indian contexts. From here, background concepts relating to socioeconomic and cultural norms can be adjusted per applicable population and tested to broaden these conclusions across other more widely used crowdfunding sites across the globe. Further, having an understanding of the factors helping make campaigns successful can potentially help inform crowdfunding platform policies. They could also help enhance transparency and accountability, as well as creating better fundraising tools and technologies, thus paving way for the betterment of future crowdfunding campaigns. 



\noindent \textbf{Our Main Insights}: Through an in-depth quantitative analysis by leveraging machine learning techniques, we found that 
1) medical campaigns make up more than 50\% of the total data, highlighting the platform's necessity to fund the treatment and recovery of beneficiaries. However, medical campaigns are the least successful campaigns in terms of achieving the goal amount; 2) Medical campaigns are negatively affected by large goal amounts, possibly contributing to the low success rates of these campaigns; 3) Attaching a large number of images has a positive impact on medical campaigns, once again highlighting the importance of establishing credibility; 4) A large volume of campaigns originate from highly populous states or major metropolitan cities in India such as Mumbai, Bangalore, Chennai, and Hyderabad, corroborating existing research on crowdfunding and demographics that access is a major factor for acceptance and usage of technology~\cite{baber2019factors}. 

\begin{table}[htb]
\centering
\begin{small}
\begin{tabular}{l l}
\toprule
\textbf{Data} & \textbf{Description} \\ 
\midrule
Number of campaigns & 119,493 \\ 
Earliest campaign date & 1st August 2019 \\ 
Latest Campaign date & 31st August 2024 \\ 
Number of cause categories & 13 \\ 
Total amount raised & \$77,359,467.52 \\ 
Total amount requested & \$1,389,136,005.32 \\ 
Number of donations & 2,295,914\\
\bottomrule
\end{tabular}
\end{small}
\caption{Overview of the campaigns.}
\label{table:dataset_description}
\end{table}

\textbf{Relevant Work}: The realm of Indian crowdfunding has witnessed a significant growth since the launch of its first set of platforms in 2012, including \emph{Ketto}, \emph{Wishberry} (now defunct) and \emph{Milaap} (\url{https://milaap.org/})\cite{suresh2020crowdfunding}. While crowdfunding platforms share a common goal, they vary in structure and typically fall into four categories: \emph{donation-based}, \emph{reward-based}, \emph{equity-based}, and \emph{debt-based}~\cite{mollick2014dynamics, massolution2013crowdfunding}. For instance, platforms like \emph{GoFundMe} (\url{https://www.gofundme.com/}) operate on a donation-based model, where donors offer support without receiving any return incentive, but \emph{Kickstarter} (\url{https://www.kickstarter.com/} and \emph{Indiegogo} (\url{https://www.indiegogo.com/}) follow a reward-based model, providing donors with either the final product or gifts as incentives. \emph{Ketto} adopts a donation-based approach (donors are not obligated to receive any incentive), but it also allows campaigners to provide personalized rewards to donors. Unlike the existing approaches, Ketto uses a `keep-it-all' approach by focusing on social causes, allowing campaigners to retain any funds raised, regardless of achieving the funding target. This prompts investigation into why people use Ketto when they have access to similar types of more popular platforms such as \emph{GoFundMe}. Previous research on Ketto has focused on specific aspects, like the impact of topical features, tax benefits and regional disparities in access to funding often examining datasets of limited size (<1000) or delve into specific platform features~\cite{muzumdar2024latent, shah2024caste, vijaya2024reconnoitering}. In contrast, our paper offers a more comprehensive analysis of the Indian crowdfunding platform using a substantially larger dataset.

\section{Data Collection and Preprocessing}

To create our dataset, we first collect campaign links using sitemaps that list all the webpages in a given domain, resulting in 121,857 campaign links. The metadata for these campaigns was then collected using the Ketto API, which offers several endpoints such as \emph{raised}, \emph{donor}, \emph{basicinfo}, \emph{comments}, and \emph{updates}, each supplying rich metadata for the campaigns. Although the campaigns' start date spans from 2016 to 2024, 99.08\% of the campaigns are from August 2019 to August 2024. Therefore, for this study, we consider campaigns from August 2019 to August 2024. From the collected data, 4,053 campaigns successfully reached their goal amount, resulting in a success rate of just 3.51\%. While most campaigners were associated with a single campaign that they would run, there were instances of multiple campaigns being run by the same campaigner, generally those associated with an NGO. Donor IDs are unique IDs given to a donor when they donate. During preprocessing, we eliminated redundant attributes---attributes with the same values for all campaigns---and those with more than 95\% values missing. We then filtered the campaigns, excluding those with campaign descriptions shorter than 19 words (the bottom 2.5\textsuperscript{th} percentile in terms of description length) since these campaigns were of poor quality when manually analyzed. Campaigns requesting less than \$23.84 (2,000 INR) were dropped since the platform requires users to have 2,000 INR as the minimum requested amount. After preprocessing, we are left with 119,493 campaigns used as the dataset (overview of this dataset is shown in Table~\ref{table:dataset_description}) for further analysis shown in our paper.



\begin{table}[htb]
\centering
\begin{small}
\begin{tabular}{l l l}
\toprule
\textbf{Statistic} & \textbf{Requested} & \textbf{Raised} \\ 
\midrule
Minimum & 23.84 & 0\\ 
Median & 3576.53 & 0 \\ 
Mean & 9,622.35& 503.28 \\ 
Maximum & 596,089.65 & 41,638.48 \\ 
Standard Deviation &  26838.08  & 2,129.70\\ 
\bottomrule
\end{tabular}
\end{small}
\caption{Summary statistics of the amount raised and goal amount (in US Dollars), computed after removing the top-120 values (0.01\% of the data) to account for potential outliers.}
\label{table:amount_stats}
\end{table}

\section{Causes and Campaigns}

India has a diverse socioeconomic landscape where the concept of crowdfunding is relatively new and younger generations are more inclined towards adopting crowdfunding~\cite{baber2019factors}. Studying and understanding how crowdfunding operates in such a unique and diverse socioeconomic landscape can uncover resource-related gaps. Many individuals resort to crowdfunding to cover different expenses due to gaps in inadequate healthcare coverage, pressing needs to address financial gaps in providing equitable resources to communities across India. Towards this goal, we leverage different statistical and natural language processing-related metrics to understand the dynamics of funds requested and raised, types of campaigns, geolocation-based insights, etc.

\begin{figure}[htb]
    \centering
    \includegraphics[width=0.9\linewidth]{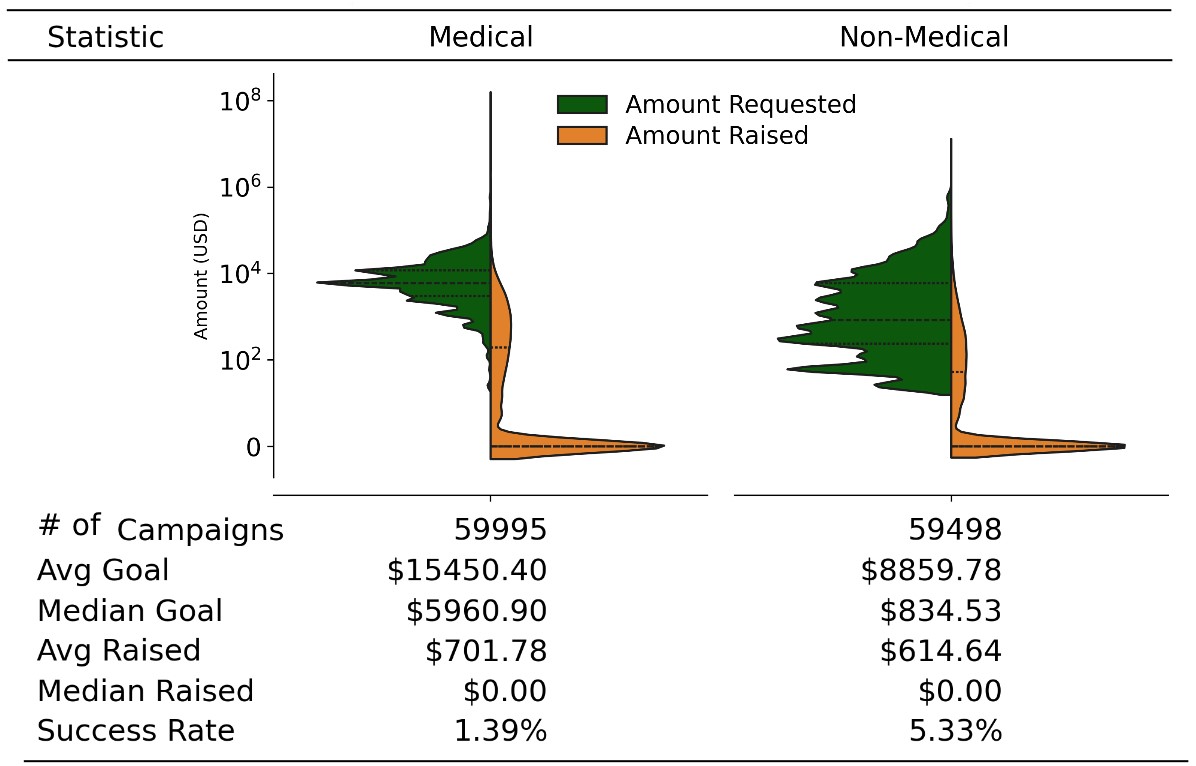}
    \caption{Distributions of requested and raised amounts for medical and non-medical campaigns followed by a detailed overview of the mean and median of the raised and requested amounts. Medical campaigns have much lower chances of succeeding than non-medical campaigns.}
    \label{fig:mednonmed}
\end{figure}

\textbf{Amounts Requested and Raised.} Due to the varying success rates across causes, we investigate how the requested and raised funds towards a particular social cause correlate in our dataset. This helps identify the ongoing needs of beneficiaries and support interest from donors but also on the trends of community engagement. Our analysis shows that the amounts requested and raised in Ketto campaigns display significant variance (as shown in Table~\ref{table:amount_stats}). Despite excluding the top 0.1\% of values in both medical and non-medical categories, the spread of requested amounts remains substantial, with an extremely high standard deviation, as seen in Table~\ref{table:amount_stats}. This high variance suggests that the financial goals of campaigns vary widely, reflecting differences in the campaign causes and their goals (Figure~\ref{fig:mednonmed}). However, campaigns initiated by NGOs have a higher probability of success and attract higher average contributions~\cite{pitschner2014non}. In our data, tax exemptions are predominantly available for campaigns run by NGOs, which also tend to focus on non-medical causes. In contrast, medical campaigns often do not offer such tax benefits. The tax benefits offered may be a critical factor in the success rate of non-medical campaigns and may also explain the reasons for low success rates for medical campaigns that were usually created by individuals. 

\begin{figure}[htb]
    \centering
    \includegraphics[width=\linewidth]{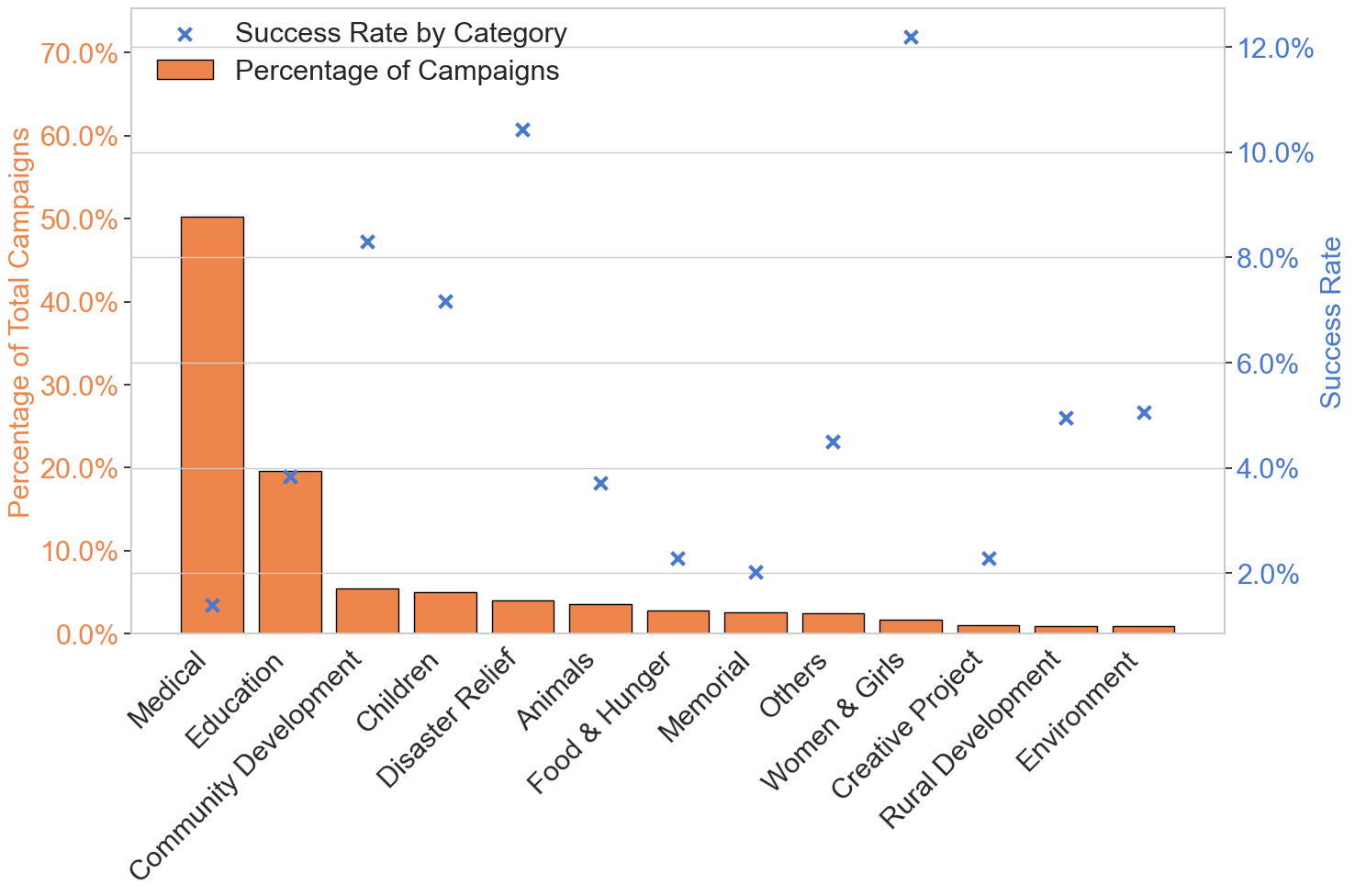}
    \caption{Distribution of campaign categories by their count and success rates. Medical campaigns make up more than 50\% of the campaigns but suffer from the least success rate.}
    \label{fig:Cause-Distribution-Success-Rates-New}
\end{figure}

\textbf{Social Causes}. Ketto allows the creation of campaigns for a variety of causes, with Medical, Education, Community Development, Children and Disaster Relief being the top 5 causes in our dataset. Medical campaigns account for more than 50\% of the dataset. In our analysis, we have considered all the campaigns associated with causes that had fewer than 1,000 campaigns grouped under the umbrella category `Others'. Despite the higher frequency of medical campaigns, as shown in Figure \ref{fig:Cause-Distribution-Success-Rates-New}, they have a notably low success rate compared to other causes.  A significant factor contributing to this disparity could be the larger donation amounts typically requested for medical campaigns, which may make it more challenging to achieve the fundraising goals. 

\begin{figure}[htb]
    \centering
    \includegraphics[width=0.8\linewidth]{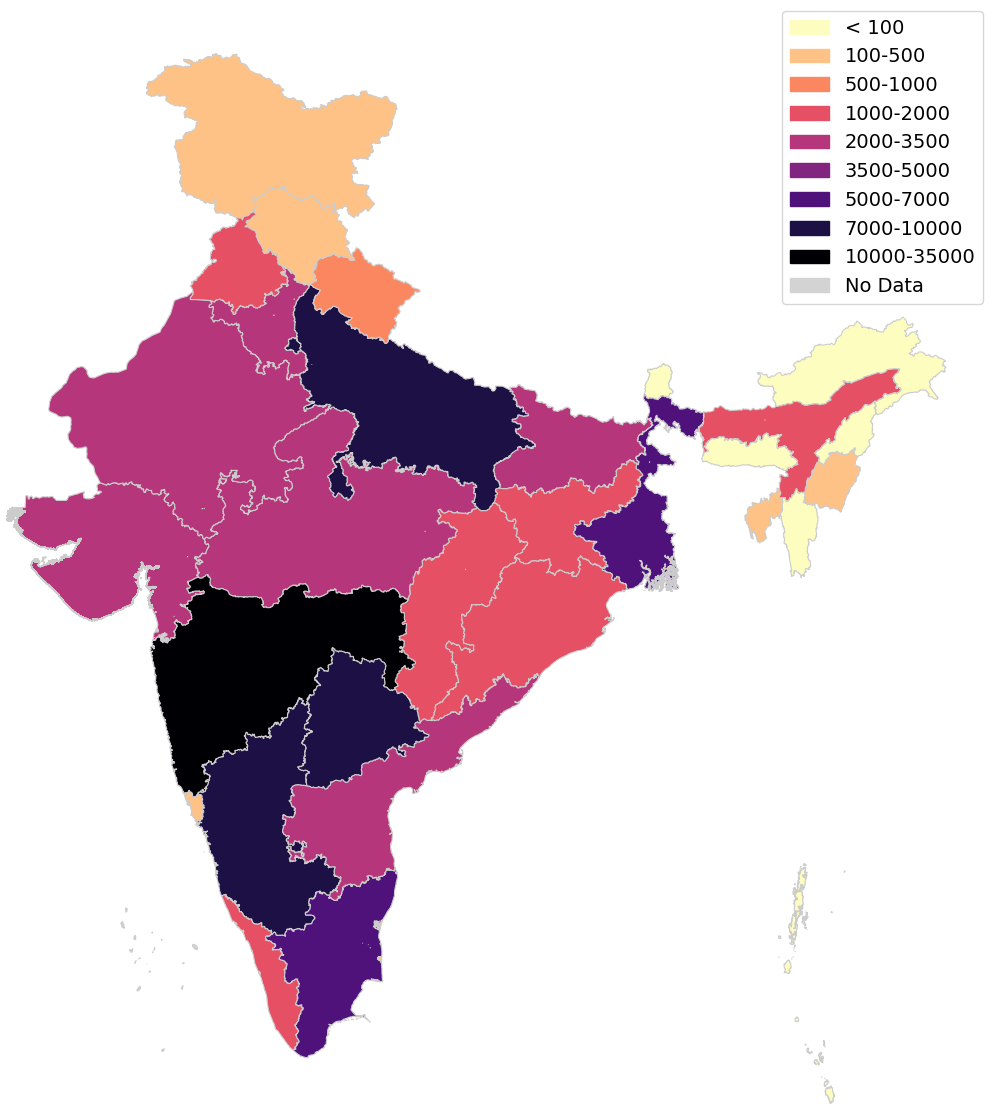}
    \caption{Number of campaigns by state with Maharashtra being the largest contributor to the number of campaigns (29.32\% of the total campaigns) originating from it.}
    \label{fig:Number-Of-Campaigns-By-Indian-State-New-Green-Discrete-Magma-cmap}
\end{figure}

\textbf{Geographical Location.} Unlike previous studies, the campaigns and beneficiaries are either primarily situated in India or from Indian subpopulations. Thus, studying geographical elements associated with campaigns is very important as it helps gain insight into \emph{access to technology}, \emph{economic trends}, \emph{cultural and social influence}, etc. A closer analysis of geolocation metadata (Figure~\ref{fig:Number-Of-Campaigns-By-Indian-State-New-Green-Discrete-Magma-cmap}) shows that Maharashtra accounts for 29.32\% of all campaigns. One reason for this could be due to the widespread presence of NGOs (13.6\% of the total active NGOs in India) in that state where, Mumbai is not only the capital city of that state but also the financial capital of India. This is followed by Karnataka (8.34\%), Uttar Pradesh (7.56\%), Telangana (6.39\%), Delhi (6.32\%), West Bengal (5.15\%), and Tamil Nadu (4.75\%)---states and union territories that are home to some of India's most populous cities. Metropolitan areas such as Mumbai, Bangalore, Chennai, Hyderabad, Delhi, and Kolkata, the state capitals, exhibit a higher concentration of individuals familiar with crowdfunding platforms. This heightened awareness in urban locations likely contributes to the significant number of campaigns originating from these regions, corroborating the existing research~\cite{brent2019economic, yu2022regional} on how crowdfunding is more likely successful in highly educated and urban areas.



\begin{figure}[htbp]
    \centering
    \includegraphics[width=\linewidth]{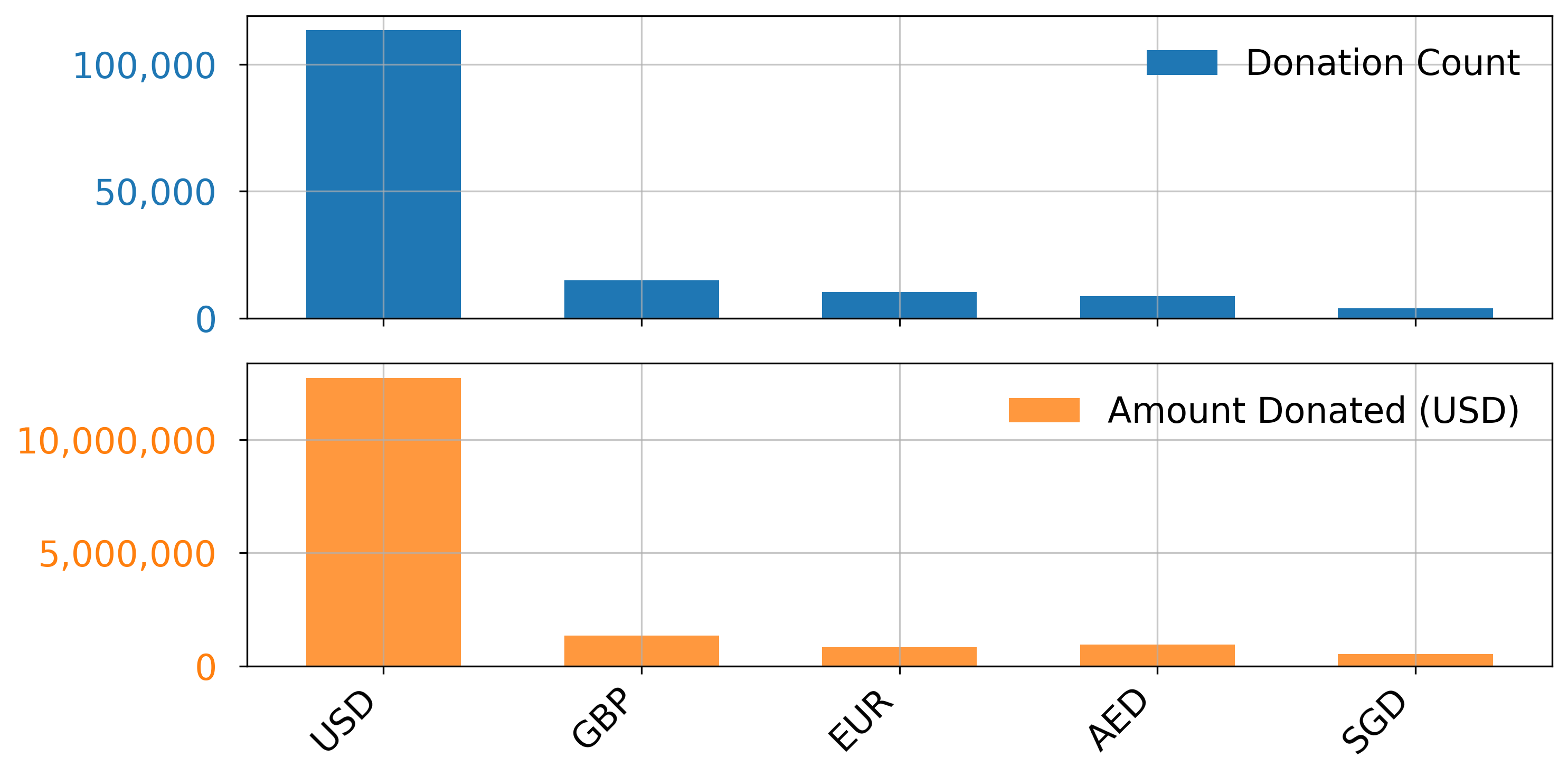}
    \caption{Number of Foreign Donations and their corresponding amounts (in USD) for the top 5 currencies.}
    \label{fig:Foreign-Donation-Count-and-Donation-Amounts-Converted-INR}
\end{figure}

\textbf{Origins of Donations.}
Ketto may primarily host campaigns in India, but a good number of donations also come from foreign countries (7.53\%), which is identified through the currency of the donation made, provided as metadata by Ketto. Some of the popular currencies other than the Indian rupees can be seen in Figure \ref{fig:Foreign-Donation-Count-and-Donation-Amounts-Converted-INR}. After Indian rupees, US dollars are the second most popular. The significant number of foreign donations coming in can be attributed to the fact that India has a worldwide diaspora and Indians in these countries could be contributing to the campaigns run in India.

\subsection{Medical Campaigns} 
From the second research question, we analyze medical campaigns to identify medical conditions for which people most frequently seek crowdfunding. Campaigns often provide data about the medical conditions for which funds are being raised. By considering all the medical conditions across the medical campaigns in our dataset, we use a keyword-based technique to prune or group synonymous terms further. For example, when there is a medical condition such as \emph{advanced liver disease} and \emph{end stage liver disease}, we count it towards the frequency of \emph{liver problems}, a more general term. Table~\ref{table:disease_count} lists the top 10 medical conditions we found out in our dataset. These conditions on Ketto represent nearly 50\% of all medical campaigns, highlighting their significant prevalence. Interestingly, while many of the leading causes of death in India are communicable diseases such as The flu, COVID-19, Tuberculosis, Malaria, etc., (please find the census data on the cause of deaths in India ~\cite{IndiaCensus}), Ketto campaigns predominantly focus on non-communicable and chronic conditions, such as cardiovascular diseases and cancer. Common ailments found both in the national mortality data and on Ketto include cancer, cardiovascular diseases, accidents, and lung-related issues.

\begin{table}[htb]
\centering
\begin{small}
\begin{tabular}{lrr}
\toprule
\textbf{Medical Condition} & \textbf{\% of medical campaigns} \\
\midrule
cancer             & 13.18 \\
heart problems      & 5.92 \\
kidney problems       & 5.91 \\
brain issues        & 5.69 \\
liver problems      & 4.97 \\
accident           & 4.56 \\
leukemia           & 3.16 \\
kidney failures     & 2.92 \\
covid-19           & 2.36 \\
lung issues         & 1.62  \\
\bottomrule
\end{tabular}
\end{small}
\caption{Top 10 most frequently featured medical conditions in our dataset of campaigns. Cancer, encompassing the different types of cancer, is one of the most prominent illness as seen by its high percentage of occurrence. }
\label{table:disease_count}
\end{table}

Cancer emerges as one of the most common reasons for fundraising on Ketto, with campaigns addressing various types, including breast, oral, lung, liver, colon, and stomach cancers, along with leukaemia and brain tumours. There is a prominence of kidney-related issues on Ketto, ranking as the third most common medical condition for which funds are sought, although they do not appear in the government census data~\cite{IndiaCensus}. \emph{These observations indicate that Ketto can serve as a valuable proxy for identifying significant medical issues in the Indian population, issues that drive individuals to seek alternate financing options like crowdfunding}. This insight could be valuable for healthcare policymakers and researchers aiming to address gaps in healthcare access and treatment affordability, especially for abundant conditions. The parallels drawn between India's government census~\cite{IndiaCensus} on leading causes of death and the data analyzed on Ketto reveal similar trends of medical issues faced by Indians. With campaigners sharing several images and documents to establish credibility and gain audience trust, hoping to lead to a donation, it is important to address the \emph{privacy concerns and issues} that might be associated with sharing personal information.

\begin{figure*}[htbp]
    \centering
    \begin{subfigure}[t]{0.475\textwidth}
        \centering
        \includegraphics[height=0.5\textwidth]{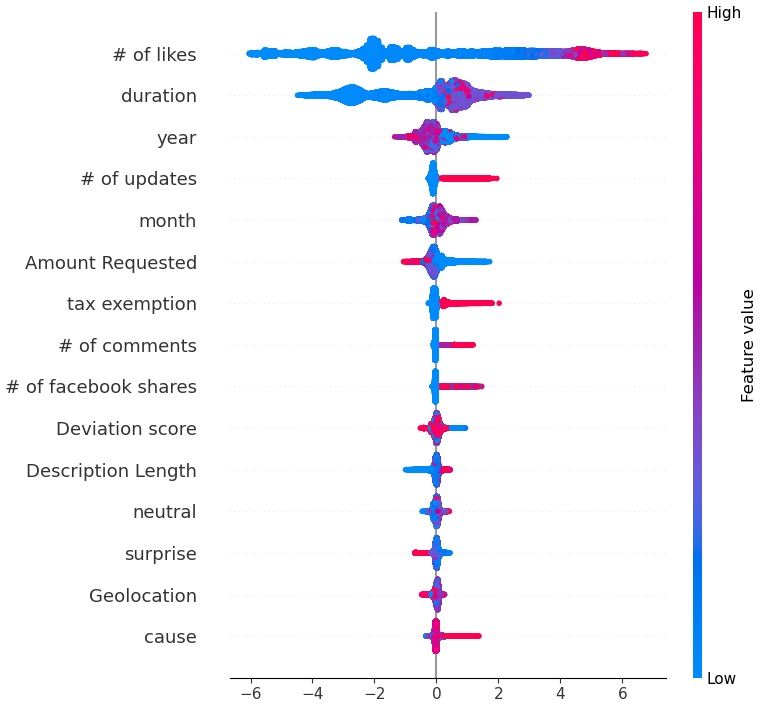}
        \caption{All campaigns}
        \label{fig:all}
    \end{subfigure}%
    ~ 
    \begin{subfigure}[t]{0.475\textwidth}
        \centering
        \includegraphics[height=0.5\textwidth]{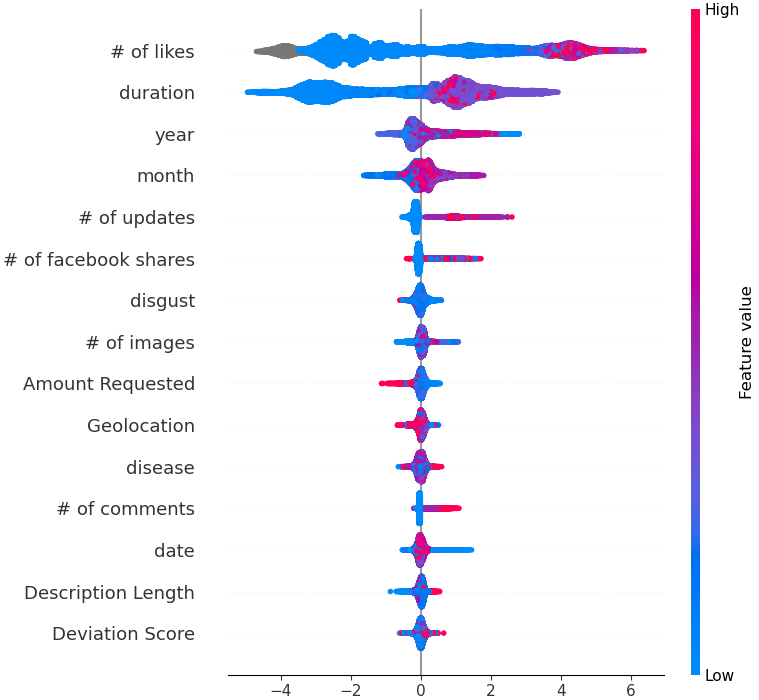}
        \caption{Medical campaigns}
        \label{fig:med}
    \end{subfigure}
    \caption{Top-15 most important features identified using SHAP}
    \label{fig:SHAP_fig}
\end{figure*}

\section{Campaign Success Factors}
With only 3.5\% of campaigns achieving their funding goals and 54.83\% of campaigns receiving no donations, understanding what leads to a successful campaign is critical, as posed in the third research question. Towards this goal, we build a classifier where we identify campaign-related features as independent attributes and consider the amount raised ($>0$ -- \emph{class-1}; otherwise -- \emph{class-2}). Key features that we extract from each campaign include user interaction features (number of comments, likes, updates, Facebook shares), number of images, campaign description length, amount requested, start date (year, month, day), campaign cause, origin location, tax benefit (1 if provided, 0 otherwise), entity type (verified NGOs vs. others), and emotions in the description (anger, disgust, fear, joy, neutral, sadness, surprise). Since there is a large percentage of medical campaigns, we built two separate models: (1) all campaigns in the dataset and (2) only medical campaigns. We use a label encoder to encode the categorical attributes and scale the data using Standard Scaler before passing it to the model. We split the data in 80:20 ratio for training and testing respectively to measure the Area Under Curve (AUC) value. We evaluated different models---XGBoost, Decision Trees, Random Forest, and Gradient Boosting, and hyperparameter---and tune the models using Optuna to maximize AUC. Since XGBoost consistently achieves the best performance, we use it for all subsequent feature importance analyses. \emph{XGBoost performs well on the data with an AUC of 0.941 on all campaigns and 0.944 for medical campaigns}. We then extract feature ranking via Shapley values separately for the two models. 

Figure \ref{fig:SHAP_fig} highlights the top 15 most important features and the spread of red dots representing the direction of their influence on campaign outcomes. For both cases, the number of likes and campaign duration are important, emphasizing that the longer a campaign remains active, the higher the probability for them to secure donations. It also shows that the higher the online engagement via likes and social media sharing, the higher the probability for a campaign to raise more money. A higher number of likes and shares suggests active audience engagement, which increases campaign visibility. The deviation score and description length are other important features, underscoring the importance of having a good campaign description that is personalized and conveys more information to catch the donors' attention. As expected, larger goal amounts negatively affect the campaigns in raising funds. 

\vspace{-0.35 cm}

\section{Conclusion and Future Work}
This preliminary study provides insights related to not just about Ketto as a platform, but also how different factors influence the success of a campaign. This is a unique and timely study compared to the existing literature showcasing that campaigns backed by NGOs or offering tax benefits serve as strong signals of legitimacy-alignment, enhancing donor trust and increasing success rates~\cite{khurana2021legitimacy}. Unique descriptions, as opposed to usage of a template, further signal authenticity and effort, fostering a personal connection between the campaigner and donor. The geographical concentration of campaigns in urban areas points to the role of shared urban identities and greater access to technology in crowdfunding adoption. As highlighted in the results, incentives like tax exemption can play a crucial role in making a campaign successful. We encourage the reader to note that the data examined in this work is biased in multiple ways. As the existing research on crowdfunding platforms from India may suggest, the campaign organizers may be young and more tech-savvy~\cite{baber2019factors} which may lead to our insights only focusing on a particular type of crowd. However, our work provides different pointers at an aggregated level in terms of how to navigate this unknown but unique socioeconomic landscape by honoring and maintaining higher standards of ethics. In the future, it would be interesting to compare and contrast this platform with other widely used crowdfunding platforms such as GoFundMe in order to identify any potential cultural confounders that are unique to these platforms.

\bibliographystyle{ACM-Reference-Format}
\bibliography{references}


\begin{thebibliography}{13}


\ifx \showCODEN    \undefined \def \showCODEN     #1{\unskip}     \fi
\ifx \showDOI      \undefined \def \showDOI       #1{#1}\fi
\ifx \showISBNx    \undefined \def \showISBNx     #1{\unskip}     \fi
\ifx \showISBNxiii \undefined \def \showISBNxiii  #1{\unskip}     \fi
\ifx \showISSN     \undefined \def \showISSN      #1{\unskip}     \fi
\ifx \showLCCN     \undefined \def \showLCCN      #1{\unskip}     \fi
\ifx \shownote     \undefined \def \shownote      #1{#1}          \fi
\ifx \showarticletitle \undefined \def \showarticletitle #1{#1}   \fi
\ifx \showURL      \undefined \def \showURL       {\relax}        \fi
\providecommand\bibfield[2]{#2}
\providecommand\bibinfo[2]{#2}
\providecommand\natexlab[1]{#1}
\providecommand\showeprint[2][]{arXiv:#2}

\bibitem[Baber(2019)]%
        {baber2019factors}
\bibfield{author}{\bibinfo{person}{Hasnan Baber}.} \bibinfo{year}{2019}\natexlab{}.
\newblock \showarticletitle{Factors underlying attitude formation towards crowdfunding in India}.
\newblock \bibinfo{journal}{\emph{International Journal of Financial Research}} (\bibinfo{year}{2019}).
\newblock


\bibitem[Brent and Lorah(2019)]%
        {brent2019economic}
\bibfield{author}{\bibinfo{person}{Daniel~A Brent} {and} \bibinfo{person}{Katie Lorah}.} \bibinfo{year}{2019}\natexlab{}.
\newblock \showarticletitle{The economic geography of civic crowdfunding}.
\newblock \bibinfo{journal}{\emph{Cities}}  \bibinfo{volume}{90} (\bibinfo{year}{2019}), \bibinfo{pages}{122--130}.
\newblock


\bibitem[Ketto(2024)]%
        {KettoWebsite}
\bibfield{author}{\bibinfo{person}{Ketto}.} \bibinfo{year}{2024}\natexlab{}.
\newblock \bibinfo{title}{Ketto: Crowdfunding India}.
\newblock \bibinfo{howpublished}{https://www.ketto.org/}.
\newblock
\newblock
\shownote{(accessed Jan 6, 2024)}.


\bibitem[Khurana(2021)]%
        {khurana2021legitimacy}
\bibfield{author}{\bibinfo{person}{Indu Khurana}.} \bibinfo{year}{2021}\natexlab{}.
\newblock \showarticletitle{Legitimacy and reciprocal altruism in donation-based crowdfunding: Evidence from India}.
\newblock \bibinfo{journal}{\emph{Journal of Risk and Financial Management}} \bibinfo{volume}{14}, \bibinfo{number}{5} (\bibinfo{year}{2021}), \bibinfo{pages}{194}.
\newblock


\bibitem[Massolution(2013)]%
        {massolution2013crowdfunding}
\bibfield{author}{\bibinfo{person}{Massolution}.} \bibinfo{year}{2013}\natexlab{}.
\newblock \bibinfo{booktitle}{\emph{2013CF: The Crowdfunding Industry Report}}.
\newblock \bibinfo{type}{{T}echnical {R}eport}. \bibinfo{institution}{Massolution}.
\newblock


\bibitem[Mollick(2014)]%
        {mollick2014dynamics}
\bibfield{author}{\bibinfo{person}{Ethan Mollick}.} \bibinfo{year}{2014}\natexlab{}.
\newblock \showarticletitle{The dynamics of crowdfunding: An exploratory study}.
\newblock \bibinfo{journal}{\emph{Journal of business venturing}} \bibinfo{volume}{29}, \bibinfo{number}{1} (\bibinfo{year}{2014}), \bibinfo{pages}{1--16}.
\newblock


\bibitem[Muzumdar et~al\mbox{.}(2024)]%
        {muzumdar2024latent}
\bibfield{author}{\bibinfo{person}{Prathamesh Muzumdar}, \bibinfo{person}{George Kurian}, {and} \bibinfo{person}{Ganga~Prasad Basyal}.} \bibinfo{year}{2024}\natexlab{}.
\newblock \showarticletitle{A Latent Dirichlet Allocation (LDA) Semantic Text Analytics Approach to Explore Topical Features in Charity Crowdfunding Campaigns}.
\newblock \bibinfo{journal}{\emph{arXiv preprint arXiv:2401.02988}} (\bibinfo{year}{2024}).
\newblock


\bibitem[Office of~the Registrar General \& Census~Commissioner(2024)]%
        {IndiaCensus}
\bibfield{author}{\bibinfo{person}{India~(ORGI) Office of~the Registrar General \& Census~Commissioner}.} \bibinfo{year}{2024}\natexlab{}.
\newblock \bibinfo{title}{India - Sample Registration System (SRS) - Cause of Death in India}.
\newblock \bibinfo{howpublished}{https://censusindia.gov.in/nada/index.php/catalog/44752}.
\newblock
\newblock
\shownote{(accessed Aug 6, 2024)}.


\bibitem[Pitschner and Pitschner-Finn(2014)]%
        {pitschner2014non}
\bibfield{author}{\bibinfo{person}{Stefan Pitschner} {and} \bibinfo{person}{Sebastian Pitschner-Finn}.} \bibinfo{year}{2014}\natexlab{}.
\newblock \showarticletitle{Non-profit differentials in crowd-based financing: Evidence from 50,000 campaigns}.
\newblock \bibinfo{journal}{\emph{Economics Letters}} \bibinfo{volume}{123}, \bibinfo{number}{3} (\bibinfo{year}{2014}), \bibinfo{pages}{391--394}.
\newblock


\bibitem[Shah(2024)]%
        {shah2024caste}
\bibfield{author}{\bibinfo{person}{Arpit Shah}.} \bibinfo{year}{2024}\natexlab{}.
\newblock \showarticletitle{Caste Inequality in Medical Crowdfunding in India}.
\newblock \bibinfo{journal}{\emph{The Journal of Development Studies}} (\bibinfo{year}{2024}), \bibinfo{pages}{1--19}.
\newblock


\bibitem[Suresh et~al\mbox{.}(2020)]%
        {suresh2020crowdfunding}
\bibfield{author}{\bibinfo{person}{Krishnamurthy Suresh}, \bibinfo{person}{Stine {\O}yna}, {and} \bibinfo{person}{Ziaul~Haque Munim}.} \bibinfo{year}{2020}\natexlab{}.
\newblock \showarticletitle{Crowdfunding prospects in new emerging markets: the cases of India and Bangladesh}.
\newblock \bibinfo{journal}{\emph{Advances in Crowdfunding: Research and Practice}} (\bibinfo{year}{2020}), \bibinfo{pages}{297--318}.
\newblock


\bibitem[Vijaya et~al\mbox{.}(2024)]%
        {vijaya2024reconnoitering}
\bibfield{author}{\bibinfo{person}{Vijaya}, \bibinfo{person}{Ajit Yadav}, {and} \bibinfo{person}{Himendu~Prakash Mathur}.} \bibinfo{year}{2024}\natexlab{}.
\newblock \showarticletitle{Reconnoitering antecedents of donation intention in donation crowdfunding campaigns: a mediating role of crowdfunding readiness}.
\newblock \bibinfo{journal}{\emph{International Review on Public and Nonprofit Marketing}} \bibinfo{volume}{21}, \bibinfo{number}{1} (\bibinfo{year}{2024}), \bibinfo{pages}{229--254}.
\newblock


\bibitem[Yu and Fleming(2022)]%
        {yu2022regional}
\bibfield{author}{\bibinfo{person}{Sandy Yu} {and} \bibinfo{person}{Lee Fleming}.} \bibinfo{year}{2022}\natexlab{}.
\newblock \showarticletitle{Regional crowdfunding and high tech entrepreneurship}.
\newblock \bibinfo{journal}{\emph{Research Policy}} \bibinfo{volume}{51}, \bibinfo{number}{9} (\bibinfo{year}{2022}), \bibinfo{pages}{104348}.
\newblock


\end{thebibliography}

\end{document}